\documentclass[10pt,letterpaper]{article}
\usepackage{opex3}
\begin{document}

\title{Probing a Bose-Einstein condensate with an atom laser}

\author{D. D\"oring, N. P. Robins, C. Figl, and J. D. Close}

\address{Australian Research Council Centre of Excellence for Quantum-Atom Optics, \\Physics Department, The Australian National University, \\Canberra, 0200, Australia}

\email{daniel.doering@anu.edu.au}

\begin{abstract} A pulsed atom laser derived from a Bose-Einstein condensate is used to probe a second target condensate. The target condensate scatters the incident atom laser pulse. From the spatial distribution of scattered atoms, one can infer important properties of the target condensate and its interaction with the probe pulse. As an example, we measure the $s$-wave scattering length that, in low energy collisions, describes the interaction between the $\left|F=1,m_F=-1\right\rangle$ and $\left|F=2,m_F=0\right\rangle$ hyperfine ground states in $^{87}\mbox{Rb}$.
\end{abstract}


\section{Introduction}

Since the realization of Bose-Einstein condensation in ultracold atomic gases, the atom laser, a highly coherent, freely propagating beam of low energy atoms has been developed by several groups \cite{M.1997, E.1999, Immanuel1999,Y.2001}. To form an atom laser, a beam of atoms is coherently output-coupled from a trapped Bose-Einstein condensate to a state that does not interact with the trapping potential. The atoms fall away from the trap producing a coherent de Broglie matter wave that is the atom laser beam. Atom lasers are the direct analogue of optical lasers. Both devices rely on Bose-enhanced scattering for their operation, and both produce coherent beams derived from a macroscopically populated trapped state. The flux, the spatial mode, the coherence properties and the quantum statistics have all been studied both experimentally and theoretically \cite{N.2006,J.-F.2006,Anton2005}. 

Despite the promise shown by the atom laser as a bright source of coherent atoms, an atom laser has not yet been used as a measurement device. In contrast, fast atomic and molecular beams derived from supersonic nozzle expansions have found widespread use in physics and physical chemistry to determine important properties such as molecular potential surfaces. In such experiments, two beams collide and scatter inside a vacuum chamber. Analyses of the angular distribution of scattering events provides information on the potentials describing the interaction between the collision partners \cite{G.1992}. This information is a basic input into many calculations in quantum chemistry. Similarly, fast atomic beams are widely used in surface science to measure properties such as the surface geometry of adsorbates, underlying surface  crystal structures and the density of states of surface phonons. In these experiments, a  fast beam strikes the surface and is scattered. The angular distribution and energy of the scattered atoms is analysed and the desired surface property is extracted \cite{G.1981, E.1992, D.1998}.  

It is an intriguing idea to explore the atom laser for analogous applications at very low collision energies. In this Letter, we present results from such an experiment. We measure the $s$-wave scattering length describing the interaction between the $\left|F=2, m_F=0\right\rangle$  and $\left|F=1, m_F=-1\right\rangle$ hyperfine ground states of $^{87}\mbox{Rb}$ by scattering an atom laser beam off a target Bose-Einstein condensate. 

The $s$-wave scattering length is an important quantity that describes atomic interactions in low energy collisions in ultracold bosonic systems. Interactions in these systems have been used to study four-wave mixing of matter waves \cite{L.1999,A.2007} and to demonstrate matter wave amplification \cite{J.M.2003a}. Low energy collisions can be exploited to produce multi-particle entanglement \cite{Olaf2003} and have potential applications in quantum computation and in producing squeezed states for precision measurements. Precise measurements of $s$-wave scattering lengths of alkali atoms have been conducted using, e.g., Raman \cite{C.2000} and photoassociative spectroscopy \cite{J.1999}. For $^{87}\mbox{Rb}$, such measurements have been accomplished in a highly accurate way. The present uncertainty in the $s$-wave scattering lengths is of the order of $0.1\,\%$ \cite{E.2002}. As rightly pointed out by Buggle \textit{et al.} \cite{Ch.2004}, these methods are based on refined knowledge of the molecular potentials. The experiment we present here does not rely on this information and offers a more direct and less complex way of measuring the interactions between two different hyperfine states in $^{87}\mbox{Rb}$. The precision of our measurement is $3\,\%$ and can most likely be improved by a more detailed theoretical analysis. In a conceptually identical setup, the technique could be adapted to study inter-species scattering and measure currently unknown interaction parameters. 

Unlike recent experiments studying the scattering properties of two colliding condensates in the same \cite{Ch.2004, Nicholas2004} or in different \cite{Angela2007} internal states, we investigate scattering in an energy regime where it is only necessary to consider $s$-wave collisions. The probing atom laser is in the $\left|F=2, m_F=0\right\rangle$ state, which is to first order magnetically insensitive and therefore reduces the impact of unstable magnetic fields on the measurement.

\section{Experimental methods}

\begin{figure}[h]
\centering \includegraphics[scale=1]{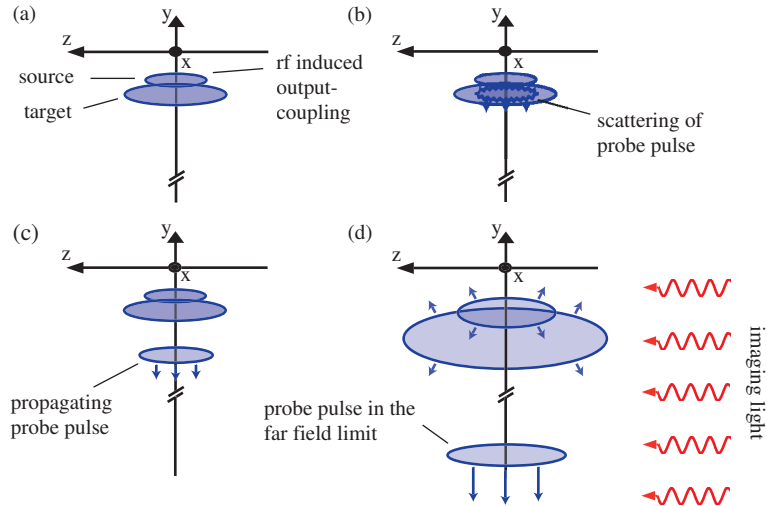}
\caption{\label{fig:scheme} (Color online) Schematic setup of the experiment. The two condensates are situated at different vertical positions in the same magnetic trap. (a) The atom laser is output-coupled from the source condensate and (b) scattered off the target condensate. (c) The scattered probe pulse falls under gravity and (d) is imaged by near-resonant light incident along the long axis of the condensates. The confining magnetic fields are switched off before the image is taken, letting the condensates expand for $5\,\mbox{ms}$.}
\end{figure}

To study the low energy atomic interactions in $^{87}\mbox{Rb}$, we derive an atom laser pulse in the $\left|F=2, m_F=0\right\rangle$ hyperfine ground state from a lasing or source condensate and collide it with a second target condensate in the $\left|F=1, m_F=-1\right\rangle$ state. A schematic diagram of the experiment is shown in Fig. \ref{fig:scheme}. The center-of-mass energy of the colliding atoms lies below $1\,\mu\mbox{K}$. Analyses of the distribution of scattered atoms allow us to determine the $s$-wave scattering length describing the interaction between atoms in these states in the low energy regime. 

\subsection{Producing two separated Bose-Einstein condensates}

Our experimental apparatus for producing two spatially separated atom clouds is based around an ultrahigh vacuum chamber operating at a pressure of $10^{-11}\,\mbox{Torr}$. In this chamber, we operate a three-dimensional magneto-optical trap (MOT) which is loaded with $10^{10}$ $^{87}\mbox{Rb}$ atoms. The loading process occurs via a cold atomic beam derived from a two-dimensional MOT. After loading, the MOT is compressed, and the confining magnetic fields are switched off. The atoms are polarization-gradient cooled in the remaining optical molasses to a temperature of $40\,\mu\mbox{K}$. We then apply a short intense laser pulse of circular polarization and resonant with the $^{87}\mbox{Rb}$ $\left|F=2\right\rangle \rightarrow \left|F'=2\right\rangle$ $\mbox{D}_2$-transition to optically pump the atoms into the $\left|F=1,m_F=-1\right\rangle$ state. By precise control of the length of this pulse (with an accuracy of $0.5\,\mu\mbox{s}$), we retain a fraction of the atoms (up to $5\times 10^9$) in the $\left|F=2,m_F=2\right\rangle$ state. Typical lengths of the optical pumping pulse are between $20\,\mu\mbox{s}$ and $40\,\mu\mbox{s}$. After preparing the internal state, a magnetic quadrupole field with a gradient of $200\,\mbox{G}/\mbox{cm}$ is switched on, confining both atoms in the $\left|1,-1\right\rangle$ and in the $\left|2,2\right\rangle$ state. By means of a mechanical translation stage, the magnetically trapped atoms are transported over a distance of $20\,\mbox{cm}$ and transferred into a harmonic magnetic trap. The temperature is further reduced by radio-frequency (rf) induced evaporative cooling. Because of the different magnetic moments of the trapped states, the cooling only works efficiently for atoms in the $\left|1,-1\right\rangle$ state. However, atoms in the more tightly confined $\left|2,2\right\rangle$ state are sympathetically cooled by elastic collisions with the $\left|1,-1\right\rangle$ atoms. Gravity shifts the potential minimum to a position vertically below the magnetic field minimum. The more tightly confined atoms in the $\left|2,2\right\rangle$ source condensate are situated above the $\left|1,-1\right\rangle$ target cloud (see Fig. \ref{fig:scheme}). The separation between the centers of the two condensates is $7.3\,\mu\mbox{m}$. The Thomas-Fermi radius of each of the clouds depends only weakly on the atom number and lies between $4\,\mu\mbox{m}$ and  $6\,\mu\mbox{m}$ (in the plane perpendicular to the long condensate axis). By turning off the repumping light during the imaging process, we verify that, before the output-coupling process, there is no cloud of atoms in the $\left|2,1\right\rangle$ state which would spatially overlap with the $\left|1,-1\right\rangle$ condensate. 

\subsection{Output-coupling, scattering and imaging the probe pulse}

Two magnetic coils are placed near the vacuum chamber, allowing us to output-couple atoms to the untrapped ($m_F=0$)-state from either condensate via rf transitions between Zeeman levels. One of the coils is used to deplete the target ($\left|1,-1\right\rangle$) condensate before the scattering experiment is carried out. We can precisely control the size of this condensate by adjusting the power of the output-coupling pulse. Due to the finite frequency width of the pulse, we cannot avoid the side effect of a partial depletion of the source ($\left|2,2\right\rangle$) condensate. We compensate for this by simultaneously adjusting the optical pumping time, so that the number of atoms in the source condensate is the same for each data point. After adjusting the number of atoms in the target condensate, the desired atom number is output-coupled from the source cloud for $80\,\mu\mbox{s}$ using the second magnetic coil \cite{comm}. The short output-coupling process allows for a high signal-to-noise ratio in the image of the probe pulse. The atoms constituting the pulse are in the $\left|2,0\right\rangle$ state. The atom laser pulse scatters as it propagates through the target condensate and is detected via absorption imaging. The circularly polarized imaging laser is resonant with the $\left|F=2\right\rangle \rightarrow \left|F'=3\right\rangle$ transition. In order to image the atoms in the $\left|F=1\right\rangle$ state, we apply a short ($1\,\mbox{ms}$) repumping pulse immediately before the image is taken. To observe an effect of the lower condensate on the propagating atom laser, it is crucial to choose the right axis for the imaging beam. The geometric factor determining the influence of the target condensate on the atom laser pulse is the gradient of the atomic density in the condensate. The highest gradient and the strongest scattering is to be expected in the plane perpendicular to the long axis of the condensate, which is therefore used as the imaging direction (see Fig. \ref{fig:scheme}(d)).

\section{Results and discussion}

The interactions between the target condensate and the probe pulse are analyzed using the spatial density distribution in the pulse for different sizes of the target condensate. We present experimental results and make a comparison to numerical simulations to get a measurement of the $s$-wave scattering length between the two states involved in the scattering process. 

\begin{figure}[h]
\centering \includegraphics[width=8.35cm]{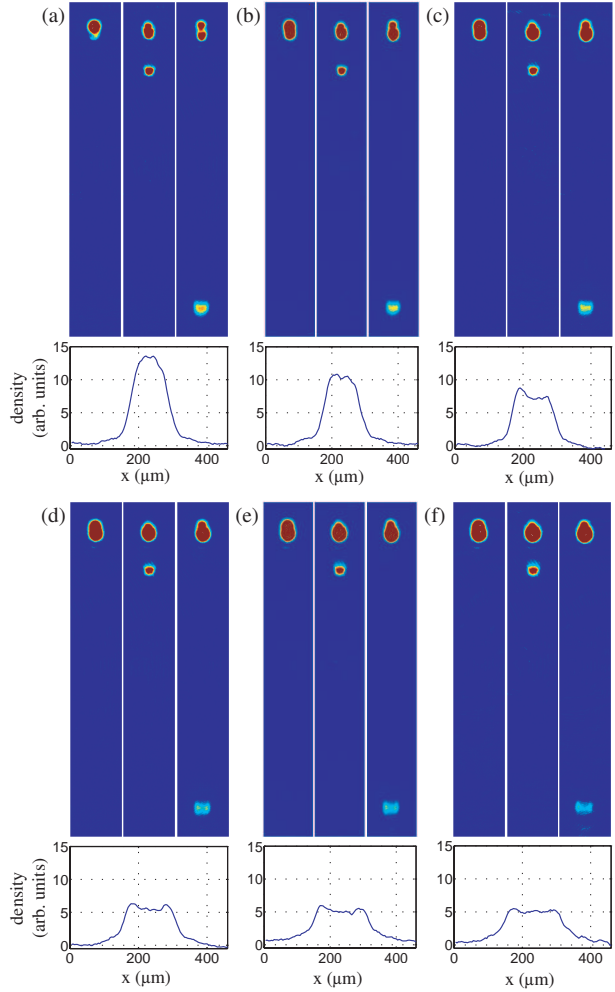}
\caption{\label{fig:results} (Color online) Non-averaged absorption pictures of the scattered probe pulse and the two condensates for atom numbers of (a) $0.05\times 10^6$, (b) $0.16\times 10^6$, (c) $0.45\times 10^6$, (d) $0.87\times 10^6$, (e) $1.25\times 10^6$, and (f) $1.38\times 10^6$ atoms in the target condensate. The three images in each section are taken $0\,\mbox{ms}$, $9\,\mbox{ms}$ and $22\,\mbox{ms}$ after output-coupling the atom laser pulse. The ballistic expansion time for the two condensates is $5\,\mbox{ms}$. The graphs below each section show the density profile of the probe pulse, integrated over the width of the pulse along the propagation ($y$-) direction.}
\end{figure}

\subsection{Experimental data}

\begin{figure}[h]
\centering \includegraphics[scale=0.58]{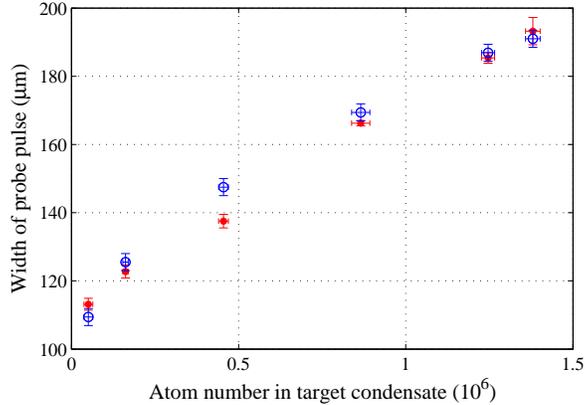}
\caption{\label{fig:graph} (Color online) Width of the scattered atom laser pulse (FWHM) as a function of the atom number in the target condensate. The red dots indicate the results of the measurements, the blue circles show the best fit of the numerical simulations to the experimental data. The fall time of the atom pulse before measuring the pulse width is $22\,\mbox{ms}$.}
\end{figure}

Increasing the size of the target condensate leads to a clear increase in the width and a change in the form of the atom laser pulse (see Fig. \ref{fig:results}). Whereas we observe the characteristic `horseshoe'-shaped pattern for the case with almost no target condensate present, this pattern changes towards a flattened profile with peaks in the atomic density on the sides of the pulse when the size of the target condensate is increased. The effect can be described by the altered momentum distribution of the probe atom laser pulse. We image the probe pulse in what can be seen as the analogue of the far field limit in classical optics. The density distribution is dominated by the momentum distribution as opposed to the change in position (which is of the order of a few $\mu\mbox{m}$) when the atom laser pulse traverses the target condensate. The effect of the altered momentum distribution on the width of the probe pulse depends proportionally on its fall time. It is crucial to optimize the fall time (to give the required spatial resolution in the image) while still maintaining a reasonable signal-to-noise ratio in the absorption image. We choose expansion times up to $22\,\mbox{ms}$. The measure used for the quantitative analysis of the scattering process is the width (FWHM) in $x$-direction of the density of the atom laser pulse, integrated along the $y$-axis (see Fig. \ref{fig:scheme}). For each set of experimental parameters, the pulse width and the atom number is averaged over five images. The measured pulse widths are shown in Fig. \ref{fig:graph}.

The collision energy of the scattering atoms is determined by the velocity that an atom has gained before it reaches the target condensate. For the fall distances considered here (the laser pulse falls less than $10\,\mu\mbox{m}$ before it reaches the target condensate), the collision energy lies below $1\,\mu\mbox{K}$. This energy is low enough to assume pure $s$-wave scattering and neglect contributions from higher order partial waves (see , e.g., \cite{Ch.2004}). 

\subsection{Numerical simulations}

The experiment allows us to extract the $s$-wave scattering length between the two states involved in the scattering process, $\left|F=1,m_F=-1\right\rangle$ and $\left|F=2,m_F=0\right\rangle$. We use a classical two-dimensional model and numerically simulate the scattering process. Quantum mechanical path interferences are neglected and can be assumed to add a fringe pattern to the classical pulse shape without significantly affecting the width of the pulse \cite{Busch2002}. The fact that we do not observe this fringe pattern in the experiment is due to the integration effect when imaging the atom laser pulse. 

In our numerical simulations, we assume the initial density profile of the atom laser pulse to be a two-dimensional Thomas-Fermi distribution, modified by a Gaussian output-coupling efficiency profile. The potential experienced by the pulse is given by Eq. (1):
\begin{equation}
U(x,y)=U_t(x,y)+U_s(x,y)+mgy \, ,
\end{equation}
where $U_t$ and $U_s$ are the potentials generated by the target and the source condensate. We obtain
\begin{equation}
U_t(x,y)= g_{12}\frac{\mu_t}{g_{11}}\left(1-\frac{(y-y_t)^2+x^2}{r_t^2}\right)\theta(r_t^2-((y-y_t)^2+x^2))\,,
\end{equation}
and an expression equivalent to Eq. (2) for the source condensate. Here, $\mu_t$ is the chemical potential, $r_t$ the Thomas-Fermi radius and $y_t$ the $y$-coordinate of the center of the target condensate. $g_{11}$ describes the coupling constant for interactions between identical atoms in the $\left|1,-1\right\rangle$ state. The inter-state coupling constant $g_{12}$ refers to scattering between atoms in the $\left|1,-1\right\rangle$ and $\left|2,0\right\rangle$ states and is related to the scattering length $a_{12}$ via $g_{12}=4\pi\hbar^2a_{12}/m$. Due to the step-function $\theta(x)$ the potential is zero outside the Thomas-Fermi radius of each condensate. 

The atom numbers of the source and the target condensate are determined from long expansion time images of the two atom clouds. The $80\,\mu\mbox{s}$ output-coupling pulse has a Fourier-limited frequency width of $16\,\mbox{kHz}$ (FWHM), and the atom laser pulse is output-coupled from a large fraction of the source condensate. There is no significant influence of the curvature of the output-coupling region, and we neglect this effect in the analysis. The number of atoms in the probe pulse is small compared to the total number of atoms in the system (the ratio is $\sim 0.05$), and we approximate the mean-field potential of both the source and the target cloud to be constant during the scattering process. The scattering length $a_{12}$ is varied in $15$ steps between $90a_0$ and $105a_0$, where $a_0$ is the Bohr radius. As the vertical position of the output-coupling region is difficult to determine accurately from the experiment, it enters the simulations as a free parameter. The best fit to the experimental data (see Fig. \ref{fig:graph}) is obtained for the center of the output-coupling region situated $2\,\mu\mbox{m}$ above the condensate center, confirming our expectation of output-coupling mainly from the central condensate region and giving a scattering length of $a_{12}=94(3)a_0$. In comparison, the values for the singlet and triplet scattering lengths in $^{87}\mbox{Rb}$ are $a_S=90.4a_0$ and $a_T=98.98a_0$ \cite{E.2002}. The stated uncertainty of $3a_0$ includes the statistical confidence region of the fit as well as the estimated systematic uncertainty due to neglecting the third dimension (along the long condensate axis) in the simulation. However, as the gradient in the atomic density along this axis is two orders of magnitude smaller than in the radial direction, the scattering in this dimension is far less distinct.

\section{Conclusion}

The work presented in this paper is the first experiment using an atom laser to probe the properties of a second independent Bose-Einstein condensate. Such a technique can take full advantage of the coherent nature of ultracold atomic samples. It is an intriguing challenge to use the coherence of a wavelength tunable (slow to fast) atom laser for the investigation of molecular potentials or for applications in surface science.

The analysis gives a measurement with an uncertainty of $3\,\%$ of the scattering length $a_{12}$ between $^{87}\mbox{Rb}$ atoms in the $\left|1,-1\right\rangle$ and $\left|2,0\right\rangle$ ground states. Theoretical analysis of the scattering of atoms on Bose-Einstein condensates has been conducted by different groups (see , e.g., \cite{Bijlsma2000a,Poulsen2003}), and we believe that a more detailed theoretical analysis of measurements like the one presented could lead to a more precise value of the scattering length $a_{12}$. The method could be extended to using a longer probe pulse or even a quasi-continuous atom laser beam. This would increase the spatial selectivity and potentially widen the applications of the conducted experiment. The use of short pulses in the work described here offers the advantage of a high signal-to-noise ratio as compared to a quasi-continuous atom laser.

Using the state-of-the-art knowledge of the singlet and triplet scattering lengths and molecular potentials, one could accurately calculate $a_{12}$. However, the method presented above does not rely on this knowledge and offers a more direct way of measuring the scattering length. In an experimental setup designed for multi-species trapping, it could be used for the study of inter-species interactions and for the measurement of currently unknown interaction parameters. 


\begin{thebibliography}{99}
\bibitem{comm}
Due to the rf ouput-coupling process we cannot avoid a weak density cloud of atoms in the $\left|2,1\right\rangle$ state. In the numerical analysis, the position of these atoms during the scattering process is approximated to be identical to the position of the source condensate.
\end{thebibliography}
\end{document}